\documentstyle[12pt]{article}
\newread\epsffilein    
\newif\ifepsffileok    
\newif\ifepsfbbfound   
\newif\ifepsfverbose   
\newdimen\epsfxsize    
\newdimen\epsfysize    
\newdimen\epsftsize    
\newdimen\epsfrsize    
\newdimen\epsftmp      
\newdimen\pspoints     
\def\epsfbox#1{\global\def\epsfllx{72}\global\def\epsflly{72}%
   \global\def\epsfurx{540}\global\def\epsfury{720}%
   \def\lbracket{[}\def\testit{#1}\ifx\testit\lbracket
   \let\next=\epsfgetlitbb\else\let\next=\epsfnormal\fi\next{#1}}%
\def\epsfgetlitbb#1#2 #3 #4 #5]#6{\epsfgrab #2 #3 #4 #5 .\\%
   \epsfsetgraph{#6}}%
\def\epsfnormal#1{\epsfgetbb{#1}\epsfsetgraph{#1}}%
\def\epsfgetbb#1{%
\openin\epsffilein=#1
\ifeof\epsffilein\errmessage{I couldn't open #1, will ignore it}\else
   {\epsffileoktrue \chardef\other=12
    \def\do##1{\catcode`##1=\other}\dospecials \catcode`\ =10
    \loop
       \read\epsffilein to \epsffileline
       \ifeof\epsffilein\epsffileokfalse\else
          \expandafter\epsfaux\epsffileline:. \\%
       \fi
   \ifepsffileok\repeat
   \ifepsfbbfound\else
    \ifepsfverbose\message{No bounding box comment in #1; using defaults}\fi\fi
   }\closein\epsffilein\fi}%
\def\epsfsetgraph#1{%
   \epsfrsize=\epsfury\pspoints
   \advance\epsfrsize by-\epsflly\pspoints
   \epsftsize=\epsfurx\pspoints
   \advance\epsftsize by-\epsfllx\pspoints
   \epsfxsize\epsfsize\epsftsize\epsfrsize
   \ifnum\epsfxsize=0 \ifnum\epsfysize=0
      \epsfxsize=\epsftsize \epsfysize=\epsfrsize
     \else\epsftmp=\epsftsize \divide\epsftmp\epsfrsize
       \epsfxsize=\epsfysize \multiply\epsfxsize\epsftmp
       \multiply\epsftmp\epsfrsize \advance\epsftsize-\epsftmp
       \epsftmp=\epsfysize
       \loop \advance\epsftsize\epsftsize \divide\epsftmp 2
       \ifnum\epsftmp>0
          \ifnum\epsftsize<\epsfrsize\else
             \advance\epsftsize-\epsfrsize \advance\epsfxsize\epsftmp \fi
       \repeat
     \fi
   \else\epsftmp=\epsfrsize \divide\epsftmp\epsftsize
     \epsfysize=\epsfxsize \multiply\epsfysize\epsftmp   
     \multiply\epsftmp\epsftsize \advance\epsfrsize-\epsftmp
     \epsftmp=\epsfxsize
     \loop \advance\epsfrsize\epsfrsize \divide\epsftmp 2
     \ifnum\epsftmp>0
        \ifnum\epsfrsize<\epsftsize\else
           \advance\epsfrsize-\epsftsize \advance\epsfysize\epsftmp \fi
     \repeat     
   \fi
   \ifepsfverbose\message{#1: width=\the\epsfxsize, height=\the\epsfysize}\fi
   \epsftmp=10\epsfxsize \divide\epsftmp\pspoints
   \vbox to\epsfysize{\vfil\hbox to\epsfxsize{%
      \includegraphics{#1}%
      \hfil}}%
\epsfxsize=0pt\epsfysize=0pt}%

{\catcode`\%=12 \global\let\epsfpercent=
\long\def\epsfaux#1#2:#3\\{\ifx#1\epsfpercent
   \def\testit{#2}\ifx\testit\epsfbblit
      \epsfgrab #3 . . . \\%
      \epsffileokfalse
      \global\epsfbbfoundtrue
   \fi\else\ifx#1\par\else\epsffileokfalse\fi\fi}%
\def\epsfgrab #1 #2 #3 #4 #5\\{%
   \global\def\epsfllx{#1}\ifx\epsfllx\empty
      \epsfgrab #2 #3 #4 #5 .\\\else
   \global\def\epsflly{#2}%
   \global\def\epsfurx{#3}\global\def\epsfury{#4}\fi}%
\def\epsfsize#1#2{\epsfxsize}
\begin{document}

\author{T. P. Singh$^* $ and Louis Witten$^\dagger $  \\ 
\\$^* $Theoretical Astrophysics Group, \\ Tata Institute of Fundamental
Research, \\ Homi Bhabha Road, Bombay 400 005 India\\ {\ e-mail:
tpsingh@tifrvax.tifr.res.in}\\ \\ $^\dagger $Department of Physics,\\
University of Cincinnati,\\ Cincinnati, OH 45221-0011, U. S. A.\\ {\ e-mail:
witten@physunc.phy.uc.edu}}
\title{Cosmic Censorship and Spherical Gravitational Collapse with Tangential
Pressure }
\maketitle
\newpage
\begin{abstract}
We study the spherical gravitational collapse of a compact object under the
approximation that the radial pressure is identically zero, and the
tangential pressure $p_{\theta }$ is related to the density $\rho $ by a
linear equation of state $p_{\theta }=k\rho $. It turns out that the
Einstein equations can be reduced to the solution of an integral for the
evolution of the area radius. We show that for positive $k$ there is a
finite region near the center which necessarily expands outwards, if
collapse begins from rest. This region could be surrounded by an inward
moving one which could collapse to a singularity - any such singularity will
necessarily be covered by a horizon. For negative $k$ the entire object
collapses inwards, but any singularities that could arise are not naked.
Thus the nature of the evolution is very different from that of dust, even
when $k$ is infinitesimally small.
\end{abstract}

\newpage

\section{Introduction}

The cosmic censorship conjecture remains one of the most important unsolved
problems in classical general relativity. In physical terms the conjecture
states that singularities arising in generic gravitational collapse are
hidden behind an event horizon. The lack of a general approach for
investigating censorship has lead to studies of simplified collapse models,
and in particular, study of spherical gravitational collapse. The best
understood spherical collapse is that of dust, where it has been shown that
for certain initial data the central singularity which forms in collapse is
naked, whereas for other initial data it is covered \cite{dust}. A more
realistic collapse model would invoke pressure. While there are general
existence arguments suggesting the occurence of naked singularities when
pressure is included \cite{dj}, explicit examples highlighting the role of
initial data are rare. It is thus not known how the naked singularity
arising in dust collapse is modified, when pressure is introduced. An
important exception is the numerical study of self-similar collapse of a
perfect fluid, where it was shown that naked singularities result
generically in such a collapse \cite{ori}.

The energy-momentum tensor for spherically collapsing matter is diagonal in
the comoving frame, the components being the energy density, the radial
pressure, and the tangential pressure. Evidently, the general analytical
solution of the Einstein equations for this system is not known, even after
an equation of state is specified. Major simplification of the system of
equations results if the radial pressure is set to zero, and only the
tangential pressure retained. This is the approximation studied in the
present paper. Admittedly, it is rather unphysical to consider a collapsing
system in which there is only tangential pressure. However, our purpose is
to explicitly study how the introduction of pressure modifies the dust
scenario, and the present analysis should be considered a small step in that
general direction.

We assume that the tangential pressure $p_{\theta}$ is related to the energy
density by a linear equation of state $p_{\theta}=k\rho$ where $k$ is a
constant. As a result, the Einstein equations can be reduced to a non-linear
ordinary differential equation for the evolution of the area radius of a
fluid element. The initial density and velocity profiles of the collapsing
object enter as free parameters in this equation, which has a one parameter
family of solutions, labelled by the parameter $k$.

The equation can be solved exactly for some specific values of $k$, and can
be integrated numerically for a general, given, $k$. However, our aim is to
check whether any singularities that form are naked, and that can be done
without resorting to numerical study. The behaviour that we find is quite
surprising, and quite unlike the evolution of dust. For positive $k$, we
find there is a region near the center which necessarily moves outwards, if
started from rest, and just cannot collapse. Depending on how far the star
extends, this outgoing region could be surrounded by a region which
collapses inwards, and could develop a singularity. However, the singularity
will not be naked, but covered by a horizon. Beyond this collapsing region,
there will in general be a second region of expansion. For negative $k$, the
entire object collapses inwards, unlike for positive $k$, but any
singularities that arise are covered and not naked. Thus the naked
singularities in dust collapse are removed when a non-zero value of $k$ is
switched on, howsoever small. Of course, this may not be the case when the
radial pressure is included, or when the tangential pressure obeys a
different equation of state. It does however suggest the possibility that
when collapse to a naked singularity occurs, the stability of the collapse
in the presence of small perturbations should be examined.

In Section 2 we describe the relevant field equations, and in Section 3 the
evolution for positive $k$ (including the case $k=1/4$ which can be solved
exactly). In Section 4 we describe the evolution for negative $k$.

Spherically symmetric static and quasi-static solutions with only tangential
pressure have been studied before, notably by Lemaitre \cite{lem}, as well
as by others \cite{oth}. Those results do not appear to have any direct
bearing on or connection with the results reported here.

\section{The Field Equations}

In comoving coordinates $(t,r,\theta ,\phi )$ the spherically symmetric
line-element is given by

\begin{equation}
\label{metric}ds^2=e^\sigma dt^2-e^\omega dr^2-R^2d\Omega ^2 
\end{equation}
where $\sigma $ and $\omega $ are functions of $t$ and $r$. The area radius $%
R$ also depends on both $t$ and $r.$ In comoving coordinates the
energy-momentum tensor for a spherically symmetric object takes the diagonal
form $T_k^i=(\rho ,p_r,p_\theta ,p_\theta )$. The quantities $p_r$ and $%
p_\theta $ are usually referred to as the radial and tangential pressure,
respectively. The Einstein field equations for this system are

\begin{equation}
\label{mprime}m^{\prime }=4\pi \rho R^2R^{\prime }, 
\end{equation}

\begin{equation}
\label{mdot}\dot m=-4\pi p_rR^2\dot R, 
\end{equation}

\begin{equation}
\label{sigpri}\sigma ^{\prime }=-\frac{2p_r^{\prime }}{\rho +p_r}+\frac{%
4R^{\prime }}{R(\rho +p_r)}(p_\theta -p_r), 
\end{equation}

\begin{equation}
\label{omedot}\dot \omega =-\frac{2\dot \rho }{\rho +p_r}-\frac{4\dot R(\rho
+p_\theta )}{R(\rho +p_r)}, 
\end{equation}
and

\begin{equation}
\label{energy}m=\frac 12R\left( 1+e^{-\sigma }\dot R^2-e^{-\omega }R^{\prime
^2}\right) . 
\end{equation}
Here, $m(t,r)$ is a free function arising out of integration of the Einstein
equations. Its initial value, $m(0,r),$ is interpreted as the mass interior
to the coordinate $r$.

We now make the approximation that the radial pressure is identically zero.
Eqn. (\ref{mprime}) remains unchanged, whereas Eqn. (\ref{mdot}) gives the
important result that the mass-function does not change with time, and
remains constant at its initial value $m(0,r)\equiv m(r)$. This is what
simplifies the analysis very considerably. Eqn. (\ref{energy}) remains as
above, while Eqns. (\ref{sigpri}) and (\ref{omedot}) simplify to

\begin{equation}
\label{sigpri2}\sigma ^{\prime }=\frac{4R^{\prime }}R\frac{p_\theta }\rho , 
\end{equation}
and

\begin{equation}
\label{omedot2}\dot \omega =-\frac{2\dot \rho }\rho -\frac{4\dot R}R\left( 1+%
\frac{p_\theta }\rho \right) . 
\end{equation}

Next we assume, for simplicity, that the equation of state is $p_\theta
=k\rho $, where the constant $k$ is chosen to lie in the range $-1\leq k\leq
1$, in accordance with the weak energy condition, and the dominant energy
condition. Eqn. (\ref{sigpri2}) then has the solution

\begin{equation}
\label{sigsol}e^{\sigma (t,r)}=R^{4k} 
\end{equation}
where a pure function of time has been set to unity. (This choice merely
reflects a rescaling of the time coordinate in the metric). Eqn. (\ref
{omedot2}) has the solution

\begin{equation}
\label{omesol}e^{-\omega (t,r)}=\psi (r)\rho ^2R^{4(1+k)}. 
\end{equation}
Here, $\psi (r)$ is an aribtrary function of the position coordinate $r$
which we retain because we would like to use this freedom later. Using Eqn. (%
\ref{mprime}) we can write this as

\begin{equation}
\label{omesol2}e^{-\omega (t,r)}=\frac{\psi (r)m^{\prime ^2}}{16\pi ^2}\frac{%
R^{4k}}{R^{\prime ^2}}\equiv C(r)\frac{R^{4k}}{R^{\prime ^2}}. 
\end{equation}
Note that $C(r)$ is a positive function. Using the solutions (\ref{sigsol})
and (\ref{omesol2}) in Eqn. (\ref{energy}) yields the following equation for
the evolution of the area radius: 
\begin{equation}
\label{reqn}\dot R^2+R^{4k}-2m(r)R^{4k-1}-C(r)R^{8k}=0. 
\end{equation}

This non-linear differential equation, which can be integrated in principle,
gives a one-parameter solution of Einstein equations. While the function $%
m(r)$ is determined by the initial density profile, the function $C(r)$
becomes known once the initial velocity profile is also given. Note that
this simplified equation for $R$ results because of the simplifying
assumptions that the radial pressure is identically zero and that the
tangential pressure obeys the linear equation of state. If $R(t,r)$ could be
solved for from this equation, the evolution of the density and the metric
components follows from the earlier equations. We note that the special case
of dust can be recovered from the above equation by setting $k=0$. That
yields the familiar equation for dust collapse,

\begin{equation}
\label{dust}\dot R^2=\frac{2m(r)}R+f(r), 
\end{equation}
where we have set $C(r)\equiv 1+f(r)$.

We shall now study the properties of Eqn. (\ref{reqn}) and its implications
for collapse and singularities, treating the cases of positive and negative $%
k$ separately.

\section{The evolution for positive k}

Evidently, Eqn. (\ref{reqn}) cannot be solved analytically for an arbitrary
value of $k$, though numerical integration for a given value of $k$ would be
straightforward. Our interest here, however, is not so much in the exact
solution of this equation, but rather in those qualitative features which
would help us determine whether singularities form in collapse, and if so,
whether they are naked. To this end, we note that for a fixed $r$ Eqn. (\ref
{reqn}) can be thought of as the equation for the one-dimensional motion of
a zero-energy particle, along coordinate $R$, in the potential $V(R,r)$
given by

\begin{equation}
\label{potn}V(R,r)=R^{4k}-2m(r)R^{4k-1}-C(r)R^{8k}. 
\end{equation}

Let us analyse the shape of this potential, as a function of $R$. By writing 
$V(R,r)$ as

\begin{equation}
\label{dub}V(R,r)=R^{4k-1}(R-2m-CR^{4k+1})\equiv R^{4k-1}W(R,r) 
\end{equation}
we note that $V(R,r)$ will be positive if and only if the function $W(R,r)$
is positive. Consider first the case of positive $k$. The extremum of $%
W(R,r) $ is at the point $R_0$ given by

\begin{equation}
\label{ext}R_0^{4k}(r)=\frac 1{C(r)(1+4k)}. 
\end{equation}
It is easily checked that this extremum is a maximum. It follows from (\ref
{dub}) that in order for $W(R_0,r)$ to be positive, the following condition
should be satisfied :

\begin{equation}
\label{max}(2m)^{4k}C<\frac{(4k)^{4k}}{(1+4k)^{(1+4k)}}. 
\end{equation}

We will assume that collapse begins from rest - it then follows from Eqn. (%
\ref{reqn}) that $C(r)$ gets fixed as

\begin{equation}
\label{cee}C(r)=\frac{1-\frac{2m}r}{r^{4k}}, 
\end{equation}
and Eqn. (\ref{max}) becomes

\begin{equation}
\label{ineq}\left( \frac{2m}r\right) ^{4k}\left( 1-\frac{2m}r\right) <\frac{%
(4k)^{4k}}{(1+4k)^{(1+4k)}}. 
\end{equation}

Note that $0\leq 2m/r\leq 1$. It is easily shown that the function $(2m/r)$
always satisfies this inequality, except when $(2m/r)=4k/(1+4k)\equiv k_{*}$%
. In the latter case, the two sides of Eqn. (\ref{ineq}) are actually equal,
and the maximum value $W(R_0,r)$ is exactly zero. It follows that the
maximum $W(R_0,r)$ is necessarily positive, except when $(2m/r)$ takes the
special value $k_{*}$. Further, since $W(R,r)$ goes to $(-2m)$ as $R$ goes
to zero, and it goes to negative infinity as $R$ goes to infinity, we
conclude that $W$ has two zeroes. These are also zeroes of the potential
function $V(R,r)$. In the region bounded by these two zeroes, $V(R,r)$ will
be positive. Thus, as can be seen from Eqn. (\ref{reqn}), this region will
be forbidden, by the requirement of positivity of $\dot R^2.$ Outside of
this forbidden region, both $W(R,r)$ and $V(R,r)$ will be negative - that is
the allowed region.

In the region to the left of the lower zero, where $V(R,r)$ is negative, the
shape of the potential depends on $k$. For $k<1/4,$ $V(R)$ monotonically
decreases as $R$ decreases, ultimately going to $-\infty $ as $R$ goes to
zero. For $k=1/4$ the potential again monotonically decreases with
decreasing $R$, going to $-2m(r)$ at $R=0$. For $k>1/4$ the potential will
have at least one turning point (which is a minimum) and goes to zero at $%
R=0 $. In the region to the right of the larger zero, where $V(R)$ is again
negative, it will go to $-\infty $ as $R\rightarrow \infty $, for all $k$.
While there are no turning points in this region for $k\geq 1/4$, there may
be turning points for $k<1/4$, depending on the functions $C(r)$ and $m(r)$.
The shape of $V(R)$ for the various cases is shown schematically in Fig. 1.

We now show that this shape of the potential function has interesting
implications for the nature of the motion. A fluid element labelled by $r$
can lie either to the right of the larger zero, or to the left of the
smaller zero. For convenience of analysis, we shall assume that at the time $%
t=0$, which marks the start of collapse, we have the scaling $R=r$. (Recall
that such a freedom of labelling is available). Since the motion starts from
rest, it is evident that the potential $V(R,r)$ is zero at the beginning.
Hence one of the two zeroes of the potential is given exactly by $R=r$. It
is important to settle whether this is the larger or smaller of the two
zeroes. For this purpose, we note that if the starting point $R=r$ lies to
the left of the maximum $R_0$, it will satisfy $r^p<R_0^p$, i.e.

\begin{equation}
\label{less}\frac{4k}{1+4k}<\frac{2m}r, 
\end{equation}
as follows after using Eqns. (\ref{ext}) and (\ref{cee}). Similarly, if the
starting point lies to the right of the maximum, it will satisfy

\begin{equation}
\label{more}\frac{4k}{1+4k}>\frac{2m}r. 
\end{equation}

\begin{center}
\leavevmode\epsfysize=4 in\epsfbox{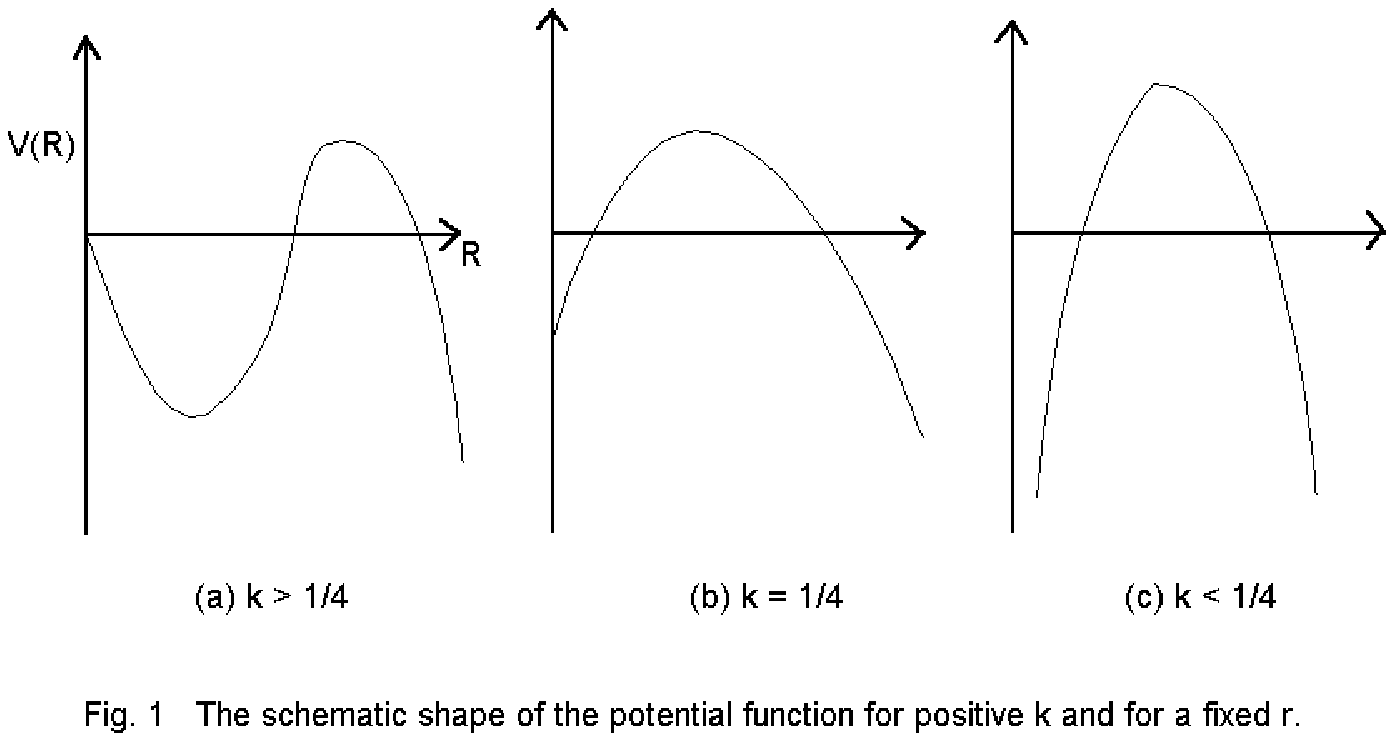}
\end{center}

Hence for all those points $r$ for which (\ref{less}) holds $R=r$ is the
smaller of the two zeroes, and all points for which (\ref{more}) holds, $R=r$
is the larger of the two zeroes. Now consider a neighborhood of the center.
The ratio $(2m/r)$ is equal to zero at the origin, and hence there will be a
neighborhood of the origin for which the inequality (\ref{more}) will
necessarily hold. All the fluid elements in this neighborhood will have
their motion begin at the larger of the two zeroes, and such elements will
necessarily move away from the origin, towards infinity. They simply cannot
collapse. Hence, there will be a neighborhood of the origin which will
necessarily move outwards, starting from rest. If the condition (\ref{more})
is satisfied all over the star, then all of it will move outwards. For $%
k\geq 1/4$ the shape of the potential implies that everything will escape to
infinity. For $k<1/4$ some or all of the fluid elements could get trapped at
a possible minimum the potential could have.

The ratio $2m(r)/r$ is zero at the origin, and near $r=0$ increases with
increasing $r$. Hence it is possible that the above-mentioned neighborhood
of the center (for which (\ref{more}) holds) is surrounded by a region in
which (\ref{less}) holds. If so, this outer region will have the left zero
of the potential as its starting point. As a result, this region will
collapse inwards, and will intersect the inner region which is moving
outwards. This appears to be a shell-crossing singularity. We will assume
here that the inner and outer regions simply cross each other. In other
words, the ingoing region does not see the outgoing inner region,
effectively.

This infalling region will necessarily shrink to $R=0$ for $k\leq 1/4$. It
could do so for $k > 1/4$ as well - i.e. it could overshoot the minimum of $%
V(R)$ for suitably chosen velocity and density profiles. $R=0$ possibly
corresponds to a curvature singularity - this would happen if the energy
density defined by Eqn. (\ref{mprime}) blows up. In order to actually check
whether this happens we have to integrate Eqn. (\ref{reqn}) to find out the
behaviour of $R^{\prime}$ as $R$ goes to zero. However, for the present, our
intention is not really to check whether singularities do form, but to show
that if they form, they are not naked.

We show this by analysing the null geodesic equation for radial geodesics.
This method has been developed in earlier papers (see, for instance \cite
{djprd}), and here we briefly recall the equation we need to use. If the
coordinate $r_0$ develops a singularity, the geodesic equation for possible
geodesics emerging from this singularity can be written as 
\begin{equation}
\label{geo}X=\lim_{r\rightarrow r_0,R\rightarrow 0}\frac{R^{\prime }}{\alpha
|r-r_0|^{(\alpha -1)}}\left( 1-\sqrt{\frac{f+2m/R}{1+f}}\right) . 
\end{equation}
The various new quantities introduced in this equation are defined as
follows. $X\equiv R/|r-r_0|^\alpha $ is defined as the tangent to a
geodesic, and the constant $\alpha $ is so chosen that the ratio $R^{\prime
}/(r-r_0)^{\alpha -1}$ is finite and non-zero in the limit of approach to
the singularity. The function $f(t,r)$ is defined using Eqn. (\ref{omesol2})
as follows: 
\begin{equation}
\label{deff}C(r)R^{4k}\equiv 1+f(t,r). 
\end{equation}
It can be shown that the singularity at $r=r_0$ will be naked if this
equation admits at least one positive real root $X=X_0$. We can easily see
that in the present case such a root will not exist. The quantity $R^{\prime
}$ is non-negative, as a result of the weak-energy condition and Eqn. (\ref
{mprime}). Note that $C(r_0)$ and $m(r_0)$ are non-zero quantities, and that
the function $f$ goes to $-1$ as the singularity is approached. As a result
the expression inside the square-root on the right hand side of the above
geodesic equation goes to $\infty $, thereby preventing the occurence of a
positive real root. Hence the singularity will not be naked.

In summary, the nature of the evolution for positive $k$ is as follows. If
the evolution starts from rest, there will decidedly be a region near the
center which will move outwards. This is the region in which the inequality (%
\ref{more}) will be satisfied. In general, this region will be surrounded by
an outer region which moves inwards, because it satisfies the inequality (%
\ref{less}). This region could become singular, but such a singularity will
not be naked. We note two further points. Firstly, the inner and outer
regions are separated by an infinitesimally thin shell where neither (\ref
{less}) nor (\ref{more}) holds, but exactly the relation $2m/r = 4k/1+4k$ is
satisfied. For this value of $r$, the two zeroes of $V(R,r)$ merge and are
equal to zero, which is also the value of the maximum of the potential. A
particle initially at rest at this value of $r$ will continue to be at rest
- this equilibrium is unstable however. The second point is that the entire
star could consist, in principle, of a succession of ingoing and ougoing
regions, depending on the ratio $2m/r$ in these regions. The generality of
the results derived above will however not be affected, even if this were
the case.

It is instructive to compare these results with those for dust collapse,
where $k=0$. For any value of $k$, howsoever small, there exists a region
near the center which will move outwards. In the limit that $k$ goes to
zero, this region shrinks to zero size. So long as such a region exists, the
value of $m(r)$ for any infalling shell is non-zero - it is this fact which
prevents the singularity from being naked. Hence collapse for $k=0$ is
qualitatively very different from what happens when $k\neq 0$.

We now illustrate our conclusions with the case $k=1/4$, which is exactly
solvable. The equation of motion (\ref{reqn}) becomes 
\begin{equation}
\label{k14}\dot R^2=CR^2+2m-R 
\end{equation}
which has the following parametric solution 
\begin{equation}
\label{soln}R=\frac{r-(4m-r)\cosh \theta }{2rC},\quad t=\frac \theta {\sqrt{C%
}}. 
\end{equation}

We assume that collapse starts from rest. The function $C(r)$ is equal to $%
(r-2m)/r^2$. The start of the evolution is at $t=0$, at which time we have $%
R=r $. The solution correctly describes evolution for all values of $r$.
Points with $r>4m$ expand outwards, the point $r=4m$ does not move, and
points with $r<4m$ move inwards. There will be a neighborhood of the center
which satisfies $r>4m$ and this region will move outwards. This is
consistent with what we expect from the inequalities (\ref{less}) and (\ref
{more}) above - for $k=1/4$ the critical dividing shell between ingoing and
outgoing regions satisfies the relation $r=4m$. If the entire star consists
only of the region satisfying $r>4m$ then all of it will expand outwards.

A singularity $R=0$ will develop for points with $r<4m$ at time $t_s(r)$
given by the singularity curve 
\begin{equation}
\label{sing}t_s(r)=\frac{\theta _s(r)}{\sqrt{C}},\quad \cosh \theta
_{s}(r)=\frac r{r-4m}. 
\end{equation}
By calculating $R^{\prime }$ we can show this is a true curvature
singularity, where the density and the Ricci scalar blow up. Further, by
applying the analysis for roots described above, it is shown that this
singularity is not naked, but covered by a horizon.

\section{The evolution for negative k}

We again analyse the shape of the potential function $V(R,r)$ and of the
function $W(R,r)$ given by Eqns. (\ref{potn}) and (\ref{dub}), this time for
negative $k$. It can be checked that as $R$ goes to infinity, $W(R,r)$ goes
to infinity. As $R$ goes to zero, $W(r)$ goes to $-2m(r)$ for $k>-1/4$, to $%
-2m-C$ for $k=-1$, and to $-\infty $ for $k<-1/4$. For $k\leq -1/4,$ $W(R)$
has no extremum, and for $k>-1/4$ it has an extremum which is a minimum.
Thus $W(R)$ will have one zero - it will be positive to the right of this
zero, and negative to its left. This point will be a zero of $V(R)$ as well,
and the potential will be positive to the right of this zero, thus giving
the forbidden region. To the left of this zero, $V(R)$ will be negative,
hence this is the allowed region. For $k\leq -1/4$ the potential
monotonically decreases to $-\infty $ as $R$ goes to zero, whereas for $%
k>-1/4$ it could have turning points (including a minimum) before going to $%
-\infty $ at $R=0$. These shapes are schematically shown in Fig. 2.

\begin{center}
\leavevmode\epsfysize=4 in\epsfbox{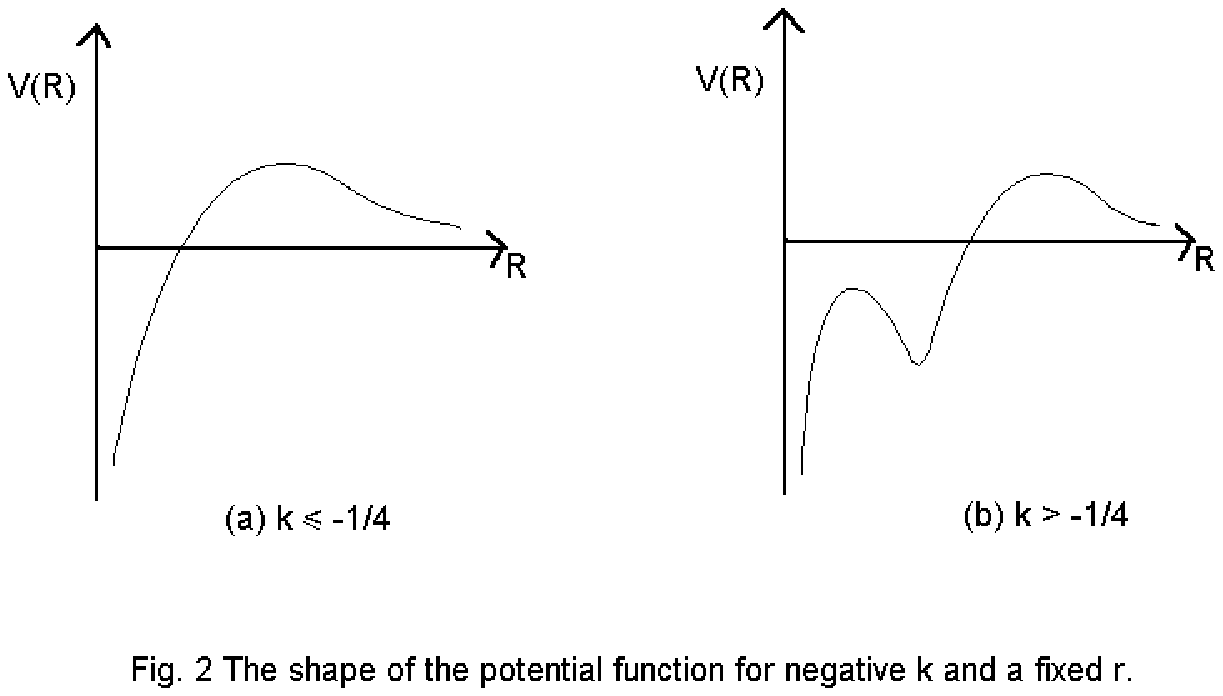}
\end{center}

Since the shape of the potential is quite different compared to when $k$ is
positive, there is no possibility of there being a region which could expand
outwards, starting from rest. The starting point of evolution is at the zero
of the potential, which is given by $R=r$. From here, the fluid element will
proceed to a smaller value of $R$, necessarily hitting $R=0$ for the case $%
k\leq -1/4$. For the case $k>-1/4$ the collapsing particles will either get
trapped at a minimum, or hit $R=0$, depending on the choice of initial
density and velocity profiles. Thus a curvature singularity could result,
for $r=0$ as well as for $r\neq 0$. We show now that such a curvature
singularity cannot be naked.

For this purpose, we again examine the geodesic equation (\ref{geo}), using
for $f(t,r)$ the solution given by Eqn. (\ref{deff}). Since $k$ is negative, 
$f(t,r)$ goes to infinity as $R$ goes to zero. Hence in the limit of
approach to the singularity, the expression inside the square-root in Eqn. (%
\ref{geo}) is equal to

\begin{equation}
\label{expn}1+\frac{2m(r)}{C(r)R^{1+4k}}. 
\end{equation}
This quantity is greater than or equal to one, which prevents the roots
equation (\ref{geo}) from having a positive real root. Hence the singularity
cannot be naked. In contrast, for the case of dust, the quantity inside the
square-root goes to infinity for $r\neq 0$, but for the singularity at $r=0$%
, this quantity is simply equal to $2m(r)/R$ (because now $f(t,r=0)=f(r=0)=0$%
). Depending on how $2m/R$ behaves in the limit, the $r=0$ singularity may
or may not be naked.

We now consider the example of $k=-1/4$, for which the equation (\ref{reqn})
for the evolution of $R$ can be explicitly solved, and the solution is

\begin{equation}
\label{negk}t=\frac 23\left( r-R\right) ^{1/2}\left( 2r+R\right) . 
\end{equation}
This solution has been written assuming that collapse begins at $t=0$ from
rest, and the scaling at $t=0$ is $R=r$. For a given shell $r$, the
singularity $R=0$ arises at the coordinate time $t_s(r)=4r^{3/2}/3$. The
quantity $R^{\prime }$ can be calculated, and is equal to $(2r-R)/R$. Hence
from Eqn.(\ref{mprime}) we get that the energy density evolves as

\begin{equation}
\label{rhoev}\rho =\frac{m^{\prime }}{4\pi R(2r-R)}. 
\end{equation}
For $r\neq 0$, $m^{\prime }$ is non-zero, implying that at $R=0$ the density
blows up, and hence there is a curvature singularity. However, the situation
is quite peculiar for $r=0$. At this point, the solution (\ref{negk}) is not
valid after $t=0$, which is just the starting epoch of collapse! Hence we
cannot talk of the density evolution at $r=0$ using this solution.

As for the points $r\neq 0$, the singularity is not naked. This is seen by
examining the roots eqn. (\ref{geo}) - the quantity inside the square root
is equal to $(r-R)/(r-2m)$, which is greater than one at the singularity $%
R=0 $. Hence there are no positive roots to this equation. An equivalent but
more physical way to arrive at this result is to work out the slope $dt/dr$
along a light ray and compare it with the slope $dt_s(r)/dr$ along the
singularity curve. If the tangent along the light ray has a larger slope
than along the singularity curve, it means the singularity lies outside the
lightcone, and hence is not naked. In the present case, the light ray is
given by

\begin{equation}
\label{light}\frac{dt}{dr}=e^{(\omega -\sigma )/2}=\frac{2r-R}{\sqrt{r-2m}},
\end{equation}
whereas along the singularity curve we have $dt_s(r)/dr=2r^{1/2}$. Thus it
follows that the slope of the light-ray is larger than that of the
singularity curve (we recall that $2m(r)/r<1$).

In this paper, we have not considered  the matching to an external
Schwarzschild solution. If the density and the constant $k$ both vanish
outside the boundary of the matter, we get the Scwarzschild solution. This
would require that $k$ actually be a function of $r$, something that we have
not presently considered. Our approach for the time being is to consider the
behaviour of solutions in regions near to the center $r=0$, where it should
be alright to let $k$ to be a constant. In principle, one will have to have
to consider matching such regions to an exterior matter region where $k$
varies with $r$, and we leave that for a future investigation.

In conclusion it could be said that in so far as tangential pressure is
concerned, the naked singularities in dust collapse are very special. They
disappear when tangential pressure is included. However, these naked
singularities will probably reappear when radial pressure is also taken into
account. Including radial pressure makes the problem analytically
untractable, as now the mass-function $m(t,r)$ evolves with time. One may
have to resort to a numerical study to tackle this problem.

The sign of the pressure appears to play a crucial role in deciding the
details of the evolution. Similar examples, where the sign of the pressure
makes an important difference in the nature of the collapse, have been noted
by Szekeres and Iyer \cite{szek}, and also by Cooperstock et al. \cite{coop}%
. Szekeres and Iyer found that in order for naked shell-focussing
singularities to occur at $r>0$, it is necessary that the radial or
tangential pressure must either be negative or equal in magnitude to the
density. Cooperstock et al. studied the non-central shell-focussing
singularity for a perfect fluid obeying the weak energy condition and showed
that with negative pressure, naked singularities can form.

Upon the completion of this work, we received a preprint by Giulio Magli 
\cite{giu} who has also examined the dynamics of a spherical matter
distribution having only tangential pressure.

\end{document}